\documentclass[aps,prb,reprint,superscriptaddress,showpacs]{revtex4-1}
\usepackage{bm}
\usepackage{graphicx}
\begin{document}

\title{Level anti-crossing spectra of nitrogen-vacancy centers in
diamond detected by using modulation of the external magnetic field}

\author{S.~V.~Anishchik}
\email[]{svan@kinetics.nsc.ru} \affiliation{Voevodsky Institute of
Chemical Kinetics and Combustion SB RAS, 630090, Novosibirsk,
Russia}

\author{K.~L.~Ivanov}
\affiliation{International Tomography Center SB RAS, 630090,
Novosibirsk, Russia} \affiliation{Novosibirsk State University,
630090, Novosibirsk, Russia}


\begin{abstract}
We report a study of the magnetic field dependence of the
photo-luminescence  of NV$^-$ centers (negatively charged
nitrogen-vacancy centers) in diamond single crystals. In such a magnetic field
dependence characteristic lines are observed, which are coming
from Level Anti-Crossings (LACs) in the coupled electron-nuclear
spin system. For enhancing the sensitivity, we used lock-in detection
to measure the photo-luminescence intensity and observed a
remarkably strong dependence of the LAC-derived lines on the
modulation frequency. Upon decreasing of the modulation frequency
from 12 kHz to 17 Hz the amplitude of the lines increases by
approximately two orders of magnitude. To take a quantitative
account for such effects, we developed a theoretical model, which
describes the spin dynamics in the coupled electron-nuclear spin
system under the action of an oscillating external magnetic field.
Good agreement between experiments and theory allows us to
conclude that the observed effects are originating from coherent spin
polarization exchange in the NV$^-$ center. Our results are of
great practical importance allowing one to optimize the
experimental conditions for probing LAC-derived lines in diamond
defect centers.
\end{abstract}


\maketitle

\section{Introduction}

The negatively charged nitrogen-vacancy center (NV$^-$ center) in
diamond  is of great interest due to its unique properties
\cite{Doherty2013}. NV$^-$ centers are promising systems for
numerous applications, in particular, for quantum information
processing
\cite{Gruber1997,Wrachtrup2001,Jelezko2004r,Childress2006,
Wrachtrup2006,Hanson2006b,Gaebel2006,Santori2006o,Waldermann2007,Maurer2012,
vanderSar2012,Dolde2013,Dolde2014,Pfaff2014} and nanoscale
magnetometry
\cite{Taylor2008,Balasubramanian2008,Maze2008,Rittweger2009,
Acosta2009,Fang2013}. It is well-known that under optical
excitation the triplet ground state of an NV$^-$ center acquires
strong electron spin polarization. Due to magnetic dipole-dipole
interactions between NV$^-$ centers and other paramagnetic defects
in the crystal spin polarization exchange can occur. Such a
polarization transfer is of relevance for many applications
\cite{Maurer2012,Jarmola2015,Mrozek2015,Chen2016}. An informative
method for studying such polarization transfer processes is given
by the Level Anti-Crossing (LAC) spectroscopy. At LACs there is no
energy barrier for polarization exchange; consequently, coupled
spins can exchange polarization. Us usual, by an LAC we mean the
following situation: at a particular field strength a pair of
levels, corresponding to quantum states $|K\rangle$ and
$|L\rangle$, tends to cross but a perturbation $V_{KL}\neq 0$
lifts the degeneracy of the levels so that the crossing is
avoided. It is well-known that at an LAC efficient coherent
exchange of populations of the $|K\rangle$ and $|L\rangle$ states
occurs
\cite{Colegrove1959,Ivanov2014,Pravdivtsev2014,Clevenson2016}.

LACs give rise to sharp lines in the magnetic field dependence of
the photo-luminescence intensity of precisely oriented NV$^-$
centers. The most prominent line \cite{Epstein2005} is observed at
1024 G, which comes from an LAC of the triplet levels in the
NV$^-$ center. Other lines are termed, perhaps, misleadingly,
cross-relaxation lines \cite{VanOort1989}. In reality, all these
lines are due to the coherent spin dynamics and are caused by spin
polarization exchange at LACs of the entire spin system of the
interacting defect centers. Thus, it is reasonable to term the
observed magnetic field dependences "LAC spectra".

In this work, we report a study of LAC-lines in diamond single crystals by using modulation
of the external magnetic field. Generally, such lines are observed by monitoring photo-luminescence as a function of the
external magnetic field; a prerequisite for such experiments
\cite{VanOort1989,Epstein2005,Hanson2006,Rogers2008,Rogers2009,Lai2009,
Armstrong2010,Anishchik2015} is precise orientation of the diamond
crystal (so that the magnetic field is parallel to [111] crystal
axis with a precision of better than one tenth of a degree).
Typically, the LAC-line at 1024~G is relatively easy to detect;
however, observing weaker satellite lines and lines coming from
interaction with other paramagnetic centers is technically more
demanding. In a previous work \cite{Anishchik2015} this problem
was minimized by using a lock-in detection: such a method provides much better sensitivity to
weaker lines. In experiments using lock-in detection the
external field strength is modulated at a frequency $f_m$; the
output luminescence signal is multiplied by the reference signal
given by $\sin(2\pi f_mt)$ (or $\cos(2\pi f_mt)$) and integrated
to provide an increased sensitivity to weak signals. In experiments
using lock-in detection \cite{Anishchik2015} a new LAC line at
zero magnetic field has been detected; groups of LAC-lines around
490-540~G, 590~G and 1024~G are clearly seen. The shape of the
lines ("dispersive" lineshape) is different from that in
conventional field-swept experiments: each line has a positive and a negative
component; at the center of each line the signal intensity is
zero. At first glance, such an appearance of the LAC-lines ("derivative" spectrum) is
standard for experiments using lock-in detection. However, here we
demonstrate an unexpected behavior of the LAC-lines, namely, a
substantial increase of the line intensity upon decrease of the
modulation frequency.

\section{Methods}
The experimental method is described in detail in a previous
publication \cite{Anishchik2015}.

Experiments were carried out using single crystals of a synthetic diamond
grown at high temperature and high pressure in a Fe-Ni-C system.
As-grown crystals were irradiated by fast electrons of an energy of
3~MeV; the irradiation dose was  $10^{18}$~el/cm$^2$. After that
the samples were annealed during two hours in vacuum at a
temperature of 800$\rm ^o$Ñ. The average concentration of NV$^-$
centers was $9.3\times10^{17}$ cm$^{-3}$.

\begin{figure}
   \includegraphics[width=0.4\textwidth]{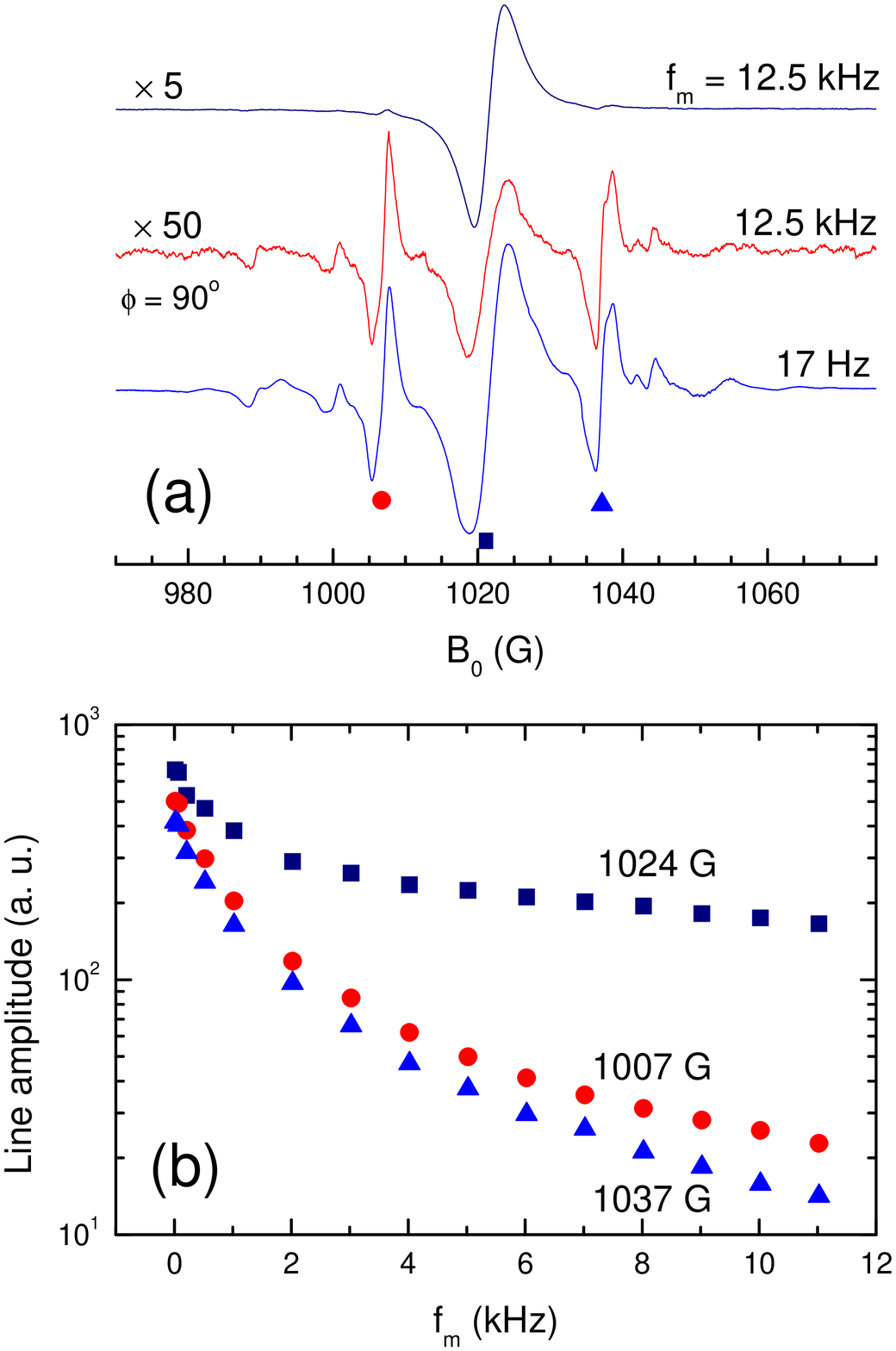}
   \caption{(a) Experimental LAC spectra of NV$^-$ centers
    in a diamond single crystal in the magnetic field range 970-1075~G.
    for each curve we give the $f_m$ value used in experiments.
    For the upper curve the phase of the lock-in detector
    is chosen such that the signal for the central LAC-line
    is maximal. For the middle trace the phase is shifted
    by $90^o$ with respect to that for the upper curve. The
    amplitude of the upper curve is increased by a factor of 5,
    for the middle curve -- by a factor of 50. The LAC-lines are indicated by circle, square and triangle. (b) Dependence
    of the amplitude of the three LAC-lines (symbols correspond to LAC-lines in subplot a) on the modulation
    frequency $f_m$. For each curve the magnetic field strength $B_0$
    corresponding to the center of the corresponding line is specified.
    For each experimental point the lock-in detector phase is set
    such that the amplitude of the corresponding line was maximal.
    In all cases the modulation amplitude was $B_m=0.5$~G. \label{experiment}}
    \end{figure}

The samples were placed in a magnetic field, which is a
superposition  of the permanent field, $B_0$, and a weak field
modulated at the frequency $f_m$:
\begin{equation}
    B=B_0+B_m \sin(2\pi f_m t), \label{bmod}
\end{equation}
and irradiated by the laser light at a wavelength of 532~nm
(irradiation power was 400 mW). The beam direction was parallel to
the magnetic field vector \textbf{B}$_0$. The laser light was
linearly polarized and the electric field vector \textbf{E} was
perpendicular to \textbf{B}$_0$. The luminescence intensity was
measured by a photo-multiplier. The resulting signal was send to
the input of the lock-in detector. The modulation frequency $f_m$
was varied from 10~Hz to 100~kHz.

\section{Results and Discussion}
Experimental results are presented in Fig. \ref{experiment}. Fig.
\ref{experiment}(a) shows the LAC-spectra of the NV$^-$ center in
the field range around 1024~G where the well-known LAC-line is
located. The spectra shown for two different modulation
frequencies, 12.5~kHz and 17~Hz, are remarkably different. As it
is seen from the Figure, when $f_m=12.5$ kHz and the lock-in
detector phase set such that the central line at 1024~G has the
maximal intensity, the satellite lines at 1007~G and 1037~G are
hardly visible. These lines are originating from polarization
exchange between the spin-polarized NV$^-$ center and neutral
nitrogen atoms (spin-1/2 defect centers), replacing carbons in the
diamond lattice \cite{Armstrong2010}. Polarization transfer occurs
when the level splittings in both defect centers become equal to
each other (causing a level crossing): under such conditions
dipole-dipole interaction turns a level crossing into an LAC and
enables coherent polarization exchange. Such a polarization
transfer is usually termed (perhaps, erroneously) cross-relaxation
\cite{VanOort1989}. The weak amplitude of the satellite lines is
due to the weak interaction between different defect centers.
Consequently, the time response of the system (in other words, the
characteristic time of the polarization transfer) is much longer
than the period of modulation, $T=1/f_m$. Therefore, there is a
phase shift of about $90^\circ$: using such a phase shift we clearly
observe the satellite lines. Interestingly, many weak additional
lines are seen when a phase shift is introduced.

When the modulation frequency is reduced to 17~Hz the amplitude of
the central line increases by a factor of 7, whereas the satellite
lines become 50 times more intense. At such a low frequency the
phase shift for all lines is very small. As it is seen from the
LAC-spectrum there are no new lines appearing in the spectrum but the
signal-to-noise ratio is substantially increased.

In Fig. \ref{experiment}(b) we present the experimental
dependences  of the line intensity, as determined for the three
different LAC-lines, on the modulation frequency. Here the total
peak-to-peak amplitude is presented; the lock-in detector phase was
set such that for each experiment the amplitude of the
corresponding line was maximal. It is clearly seen that by varying
the modulation frequency we obtained a strong variation of the
LAC-line intensities by roughly two orders of magnitude.
Interestingly, in the assessed frequency range the dependence in
non-exponential; the slope of the curve increases at lower
modulation frequencies.

The most unexpected effect is that the increase of the line
intensity is occurring at modulation frequencies, which are much
smaller than the relaxation rates of the NV$^-$ center (when
measured in the same units). To rationalize this effect we made
attempts to simulate the spin dynamics of a two-level system
having an LAC. However, such a model predicts a much smaller
effect of $f_m$; moreover, the frequency range where the line
intensity strongly varies is completely different from that found
experimentally.

The observed strong dependence can be explained by polarization transfer
from the electronic triplet spin system to nuclear spins, since
the relaxation times of nuclear spins are much longer (by several
orders of magnitude) that the electronic spin relaxation times.

To model electron-nuclear polarization transfer we made the
following simplifications. First, we did not treat the entire
three-level electron spin system but restricted ourselves to only
two levels. In such a situation the electronic spin subsystem can
be modeled by a fictitious spin $S=1/2$. We also assume that the
luminescence intensity is proportional to the population of the
$S_z=1/2$ state. This is a reasonable assumption because only one of the three triplet states of the NV$^-$ center gives intense luminescence. Hereafter we assume that the $z$-axis is parallel
to the external magnetic field. The $S$ spin interacts with the
permanent external $B_0$ field, with the oscillating $B_m$ field and with
surrounding nuclear spins by hyperfine coupling (HFC), which is
assumed to be isotropic. For the sake of simplicity, we reduced the
nuclear spin subsystem to only one spin $I=1/2$. Then the
Hamiltonian of the spin system under consideration takes the form (in $\hbar$ units):

\begin{equation}\label{Ham}
    \hat{H}(t)=\gamma B_0 \hat{S}_z + V \hat{S}_x + A (\hat{\bf S}\cdot\hat{\bf I})+\Omega_1\cos( 2\pi f_mt)\hat{S}_z,
\end{equation}
where $\hat{\bm S}$ and $\hat{\bm I}$ are the spin operators of
the electron and the nucleus, respectively, $\gamma$ is the
electronic gyromagnetic ratio, $B_0$ is the external magnetic
field strength, $V$ is an external perturbation (coming, e.g.,
from a small misalignment of the crystal), $A$ is the isotropic
HFC constant, $\Omega_1$ is the amplitude of the modulated
magnetic field, $f_m$ is the modulation frequency. Hereafter we
use the notation $\gamma B_0=\omega_0$. Considering only the main
part of the Hamiltonian, $H_0=\gamma B_0\hat{S}_z$ we obtain that
there is a level crossing at $B_0=0$; however, the perturbations
given by $V$ and $A$ mix the crossing levels and turn this
crossing into an LAC. By turning on the modulation we introduce
repeated passages through the LAC; upon these passages spin
evolution is taking place resulting in redistribution of
polarization.

Qualitatively, we expect different regimes for spin dynamics  at
low and at high $f_m$. At a low $f_m$ each passage through the LAC
results in adiabatic inversion of populations of the $S$-spin
states. Consequently, the luminescence signal is expected to be
modulated at the $f_m$ frequency having the maximal possible amplitude and the
same phase as the modulated external field. During a fast passage
through the LAC, i.e., at large $f_m$, the populations are mixed
only slightly and the amplitude of modulation of the luminescence
signal is expected to drop down. In addition, the modulation of the
signal is no longer in-phase with the reference signal of the
lock-in amplifier, resulting in strong phase shifts of the
signal. As we show below, the calculation results are in good
agreement with these expectations.

The spin evolution is described by the Liouville-von Neumann equation:
\begin{equation}\label{Liu1}
    \frac{d\rho_{ij}}{dt}=L_{ij;kl}(t)\rho_{kl},
\end{equation}
where $\rho$ is the density matrix of the two-spin
electron-nuclear  system in the Liouville representation
(column-vector with 16 elements), while the elements of the
$\hat{\hat{ L}}$ super-operator are as follows:
\begin{equation}\label{Liu2}
    L_{ij;kl}=i(\delta_{ik}H_{lj}-\delta_{jl}H_{ik})+R_{ij;kl},
\end{equation}
where $R_{ij;kl}$ is the relaxation matrix. To specify the $R$
super-operator  we made the following simplifying assumptions. We
treated two contributions to the electron spin relaxation, the
spin-lattice relaxation  (at a rate $\rm R_1$) and phase
relaxation (at a rate $\rm R_2$). In addition, we take into
account photo-excitation of the NV$^-$ center, which produces the
electron spin polarization, i.e., the population difference for the
states of the $S$-spin. This process is considered in a simplified
manner as transitions from the $S_z=-1/2$ state ($\alpha$-state)
to the $S_z=+1/2$ state ($\beta$-state) at a rate $I$
(pumping rate for the electron spin polarization). Relaxation of
the nuclear spin is completely neglected because it is usually
much slower than that for the electron spin.

The period $T=1/f_m$ was split into $N$ equal intervals of a duration $\Delta
t=T/N$. In each step, the density matrix $\rho$ was propagated by using a matrix exponent:
\begin{equation}\label{Liu3}
    \rho(t+\Delta t)=\exp[\hat{\hat{L}}(t)\Delta t]\rho(t).
\end{equation}
Generally, the solution depends on the initial conditions. However,
in  the present case we are interested in the "steady-state"
solution, which is reached after many modulation periods. Indeed,
in experiments transient effects are not important because
signal averaging is performed over many $T$ periods (only the
steady-state of the system is probed). To obtain such a solution
of eq. (\ref{Liu1}) we assume that the density matrix before a
period of modulation, $\rho(0)$, is the same as that after the
period:
\begin{eqnarray}
\nonumber  \rho(0)=\rho(T) &=& \exp[\hat{\hat{L}}(t=T-\Delta t)\Delta t]\times\ldots \\
  ~ &~& \times\exp[\hat{\hat{L}}(t=0)\Delta t]\rho(0)=\hat{\hat{U}}\rho(0).
  \end{eqnarray}
Here $\hat{\hat{U}}$ is the super-matrix, which describes the evolution over a
single modulation period. This equation is a linear equation for
the $\rho(0)$ vector. To find a non-trivial solution of such a
matrix equation, we need to exclude one equation from the system
(the one, which linearly depends on other equations) and to
replace it by the following one: $\sum_i\rho_{ii}(0)=1$, which
describes nothing else but conservation of the trace of the
density matrix. This new system can be solved by using standard
linear algebra methods. The $\hat{\hat{U}}$ matrix was computed
numerically; to do so we set the value of $N$ such that further
increase of $N$ changed the final results by less than 1\%. Of
course, it is necessary to increase $N$ substantially at small
$f_m$. At the lowest modulation frequency we used $N=4\times 10^5$.

To compare theoretical results to the experimental data we
numerically computed the sine and cosine Fourier components of
the element of interest of the density matrix, namely, the
population of the $\alpha$-state, $\rho_{\alpha\alpha}$. This
element can be computed when $\rho(t)$ is known:

\begin{equation}\label{XY1}
    X=\frac{1}{T}\int_0^{T}\rho_{\alpha\alpha}\cos(2\pi f_mt)dt,
\end{equation}

\begin{equation}\label{XY2}
    Y=\frac{1}{T}\int_0^{T}\rho_{\alpha\alpha}\sin(2\pi f_mt)dt.
\end{equation}

So, essentially, we calculate polarization of the $S$-spin, which
is given by the difference $[\rho_{\alpha\alpha}-1/2]$. In the
same way we can compute polarization of the $I$-spin. Knowing $X$ and $Y$ we can completely characterize the signal. An analogue
of the lock-in detector phase variation by an angle $\phi$ is the
rotation of axes in the functional space:
    \begin{equation}\label{newx}
    X'=X\cos\phi + Y\sin\phi.
    \end{equation}

\begin{figure}
   \includegraphics[width=0.42\textwidth]{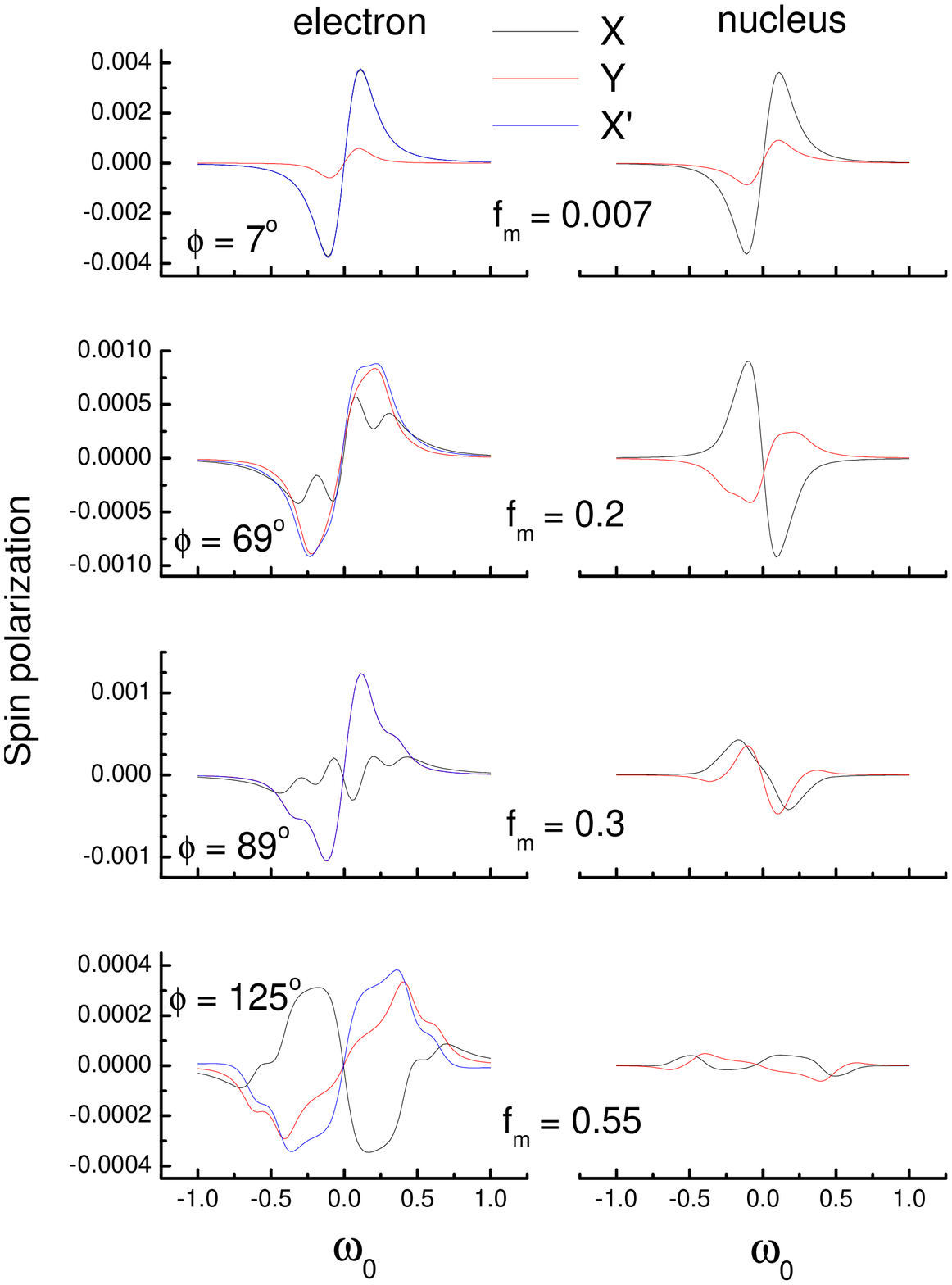}
   \caption{Theoretical LAC-spectra for different $f_m$ frequencies.
    Calculation parameters: $\rm R_1=0.1$, $\rm R_2=0.1$, $I=0.01$, $\rm V=0.1$,
    $\rm A=0.2$, $\rm \Omega_1=0.1$. The $f_m$ value is given for each curve.
    The values of all parameter are given in arbitrary frequency units.
    \label{nuclspectr}}
    \end{figure}

In Fig. \ref{nuclspectr} we present the magnetic field dependences of the Fourier $X$ and $Y$ components of the electron and nuclear polarization. Additionally, for the electron spin the $X'$ component is presented at specific values of the lock-in detector phase $\phi$. The $\phi$ phase was set such that the line intensity was maximal. One can readily see that at the lowest modulation frequency the LAC-spectra for the electron and nuclear spins are the same, resembling a dispersive Lorentzian line. At higher $f_m$ frequencies the line shape is significantly distorted and line intensity decreases. Interestingly, polarization of the nuclear spin drops faster than that for the electron spin.

\begin{figure}
   \includegraphics[width=0.42\textwidth]{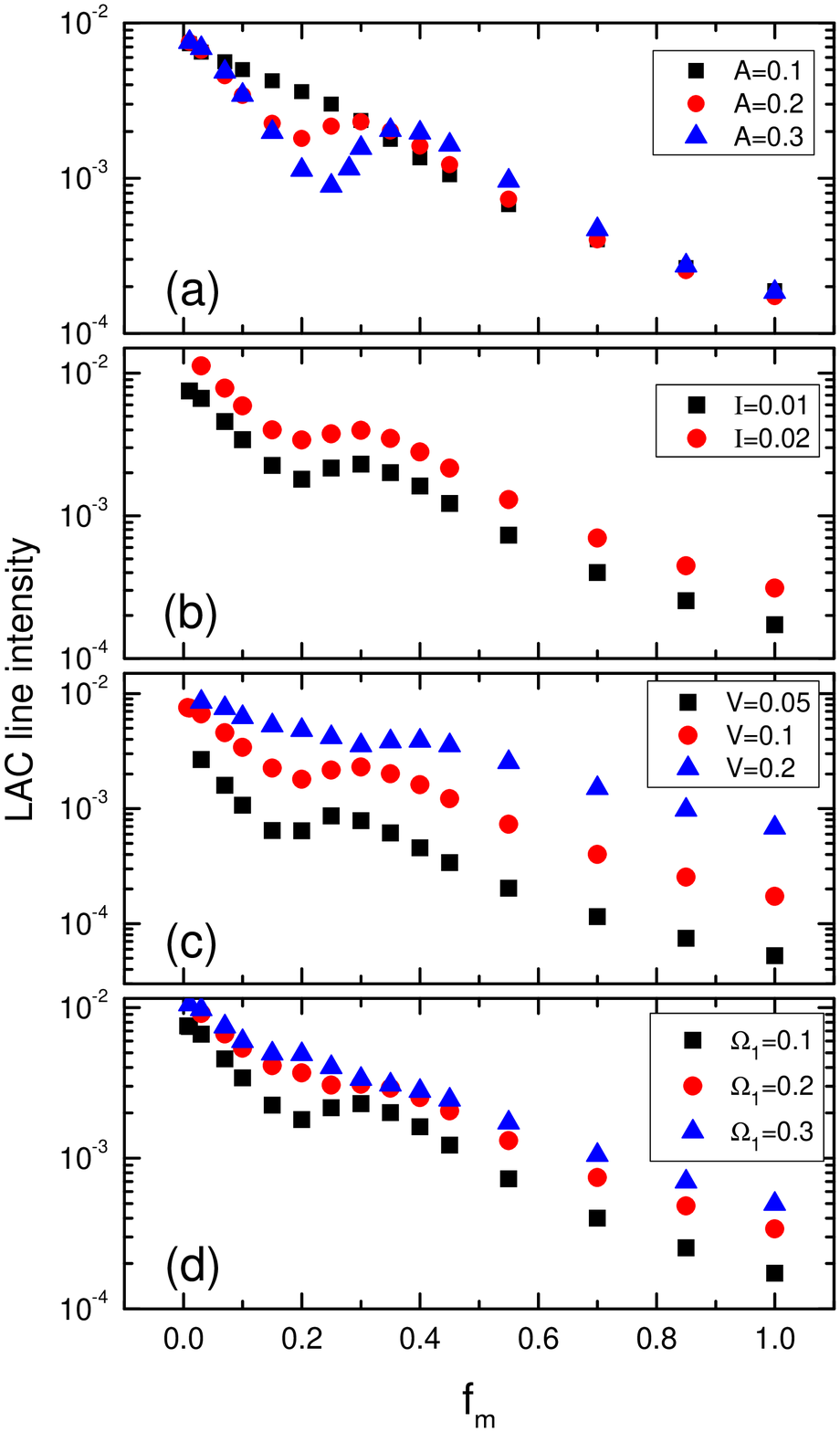}
   \caption{Calculated dependence of the LAC-line intensity on the modulation frequency $f_m$. Here we show the signal components $X$ and $Y$, as well as $X'$.
   Calculation parameters (when different from those given in the figure): $\rm R_1=0.1$, $\rm
   R_2=0.1$, $I=0.01$, $\rm V=0.1$, $\rm A=0.2$, $\rm \Omega_1=0.1$.
   The values of all parameter are given in arbitrary frequency units.\label{nucleus}}
    \end{figure}

In Fig. \ref{nucleus} we present the calculated LAC-line
intensity as functions of the modulation frequency $f_m$; the
dependences are calculated upon systematic variation of the relevant
parameters. In all cases, upon decrease of the modulation
frequency by two orders of magnitude, i.e., from 1 to 0.01, the
line intensity increases by roughly two orders of magnitude. It
also turned out that the external perturbation is of great
importance: at $\rm V\rightarrow 0$ the line intensity also tends
to zero. The line intensity also drops down upon decrease of the  $I$ parameter, simply because of the lower efficiency of polarization
formation. Likewise, the LAC-line intensity s reduced upon decrease of the
modulation amplitude, $\rm \Omega_1$; interestingly, there is no
simple proportionality relation between the line intensity and $\Omega_1$.
The dependence on the HFC constant, $\rm A$, is relatively
complicated. Notably, there is a non-monotonous dependence of the
LAC-line intensity on the modulation frequency. A plausible
explanation of this effect can be deduced from the spectra shown
in Fig. \ref{nuclspectr}. It is clearly seen that the width for
different Fourier components is considerably different: at high
frequencies the sine-component prevails, which is phase-shifted
by $\rm 90^o$ with respect to the modulation fields, whereas at
low frequencies the cosine-component is dominating. At
intermediate frequencies both components are of a comparable intensity. As a
consequence, the phase shift (required to obtain the highest
LAC-line intensity) leads to a distorted LAC-line with two
components partly compensating each other. As a result, the total
line intensity decreases.

Interestingly, the calculation predicts an even stronger  increase
of the line intensity at low $f_m$ frequencies as compared to the experimental data. Furthermore, the
theoretical dependence is closer to an exponential behavior (in
contrast to the experimental dependence). For further increase of
the LAC-line intensity at low frequencies it might be necessary to
consider an additional mechanism with even slower response. Such a
mechanism \cite{Siyushev2013} could be photo-excitation of the
excited state on an NV$^-$ center to the conduction band,
accompanied by the formation of the neutral NV$^0$ center, which
preserves polarization of the nuclear spin. When the backward
process (the electron goes to the NV$^0$ center) takes place, the
newly formed NV$^-$ has no electron spin polarization. However,
such polarization can be generated by polarization transfer from
the nuclear spin.

\section{Conclusions}
We report a study of LAC-lines in NV$^-$ defect centers in
diamond crystals by using lock-in detection of the signal. Such a
method allows one to obtain sharp LAC-lines with excellent
signal-to-noise ratio. A strong and unexpected effect of the
modulation frequency on the LAC-line intensity is demonstrated.
Interestingly, the LAC line is the strongest at low modulation
frequencies. Thus, measurements at low $f_m$ are advantageous,
even despite the technical issues concerning experiments at low
frequencies (namely, the instrumental noise). Moreover,
LAC-spectra obtained at low modulation frequencies are free from
distortions and phase shifts of the signal with respect to the
reference signal of the lock-in amplifier.

To rationalize the observed effect of the modulation  frequency we
performed a theoretical study and computed numerically the
evolution of the spin system under the action of the modulation
field. In the theoretical model, we introduced a single electron
spin 1/2 (modeling the electron spin degrees of freedom of the
NV$^-$ center) coupled to a nuclear spin 1/2. Such a model can
reproduce the main features found in experiments.
Specifically, at low modulation frequency we obtain adiabatic
exchange of populations of the states having an LAC. This results
in the modulated luminescence signal of the maximal amplitude, which
has the same phase as the external magnetic field $\Omega_1$. In
contrast, at high modulation frequencies spin mixing occurs upon
fast passages through the LAC and populations are mixed only
slightly. Consequently, the signal drops and becomes distorted. We
anticipate that these features can be reproduced also by more
elaborate calculations for extended spin systems, which can model,
e.g., polarization exchange between different paramagnetic centers
in the crystal.

Our work provides useful practical recommendations on how to
conduct experimental studies of LAC-lines. As we show is a
subsequent publications, the experimental method used here indeed
enables sensitive detection of LAC-lines and observation of
previously unknown LAC-lines. Furthermore, for the first time we
demonstrate that modulation (used in lock-in detection) is not
only a prerequisite for sensitive detection of weak signals but also a method
to affect spin dynamics of NV$^-$ centers in diamonds.

\begin{acknowledgments}

Experimental work was supported by the Russian Foundation for
Basic Research (Grant No. 16-03-00672); theoretical work was
supported by the Russian Science Foundation (grant No.
15-13-20035).

\end{acknowledgments}

\bibliography{LAC}

\end{document}